\documentclass{ws-procs9x6}
\usepackage{cite}

\newcommand{\dif}{\mathrm{d}}
\newcommand{\diff}[1]{\frac{\mathrm{d}#1}{#1}}
\newcommand{\Pom}{{\mathbb{P}}}
\newcommand{\xPom}{x_\Pom}
\newcommand{\xB}{x_{\scriptscriptstyle{B}}}
\newcommand{\chisq}{\chi^2/\mathrm{d.o.f.}}

\begin{document}

\title{Diffractive Parton Density Functions\footnote{\uppercase{T}o appear in the proceedings of the \uppercase{R}ingberg \uppercase{W}orkshop on ``\uppercase{N}ew \uppercase{T}rends in \uppercase{HERA} \uppercase{P}hysics 2005'', \uppercase{R}ingberg \uppercase{C}astle, \uppercase{T}egernsee, \uppercase{G}ermany, 2--7 \uppercase{O}ctober 2005.}}

\author{G.~WATT\footnote{\uppercase{I}n collaboration with \uppercase{A}.\uppercase{D}.~\uppercase{M}artin and \uppercase{M}.\uppercase{G}.~\uppercase{R}yskin.}}

\address{Deutsches Elektronen-Synchrotron DESY, \\
  Notkestrasse 85, 22607 Hamburg, Germany \\
  E-mail: graeme.watt@desy.de}

\maketitle

\abstracts{
  We discuss the perturbative QCD description of diffractive deep-inelastic scattering, and extract diffractive parton distributions from recent HERA data.  The asymptotic collinear factorisation theorem has important modifications in the sub-asymptotic HERA regime.  In addition to the usual \emph{resolved} Pomeron contribution, the \emph{direct} interaction of the Pomeron must also be accounted for.  The diffractive parton distributions are shown to satisfy an \emph{inhomogeneous} evolution equation, analogous to the parton distributions of the photon.
}

\section{Introduction}
A notable feature of deep-inelastic scattering is the existence of diffractive events, $\gamma^* p\to X + p$, in which the slightly deflected proton and the cluster $X$ of outgoing hadrons are well-separated in rapidity.  At high energies, the large rapidity gap is believed to be associated with `Pomeron', or vacuum quantum number, exchange.  The diffractive events make up an appreciable fraction of all (inclusive) deep-inelastic events, $\gamma^* p \to X$.  We will refer to the diffractive and inclusive processes as DDIS and DIS respectively.  The recent improvement in the precision of the DDIS data \cite{H1data,Chekanov:2004hy,Chekanov:2005vv} allow improved analyses to be performed and more reliable diffractive parton density functions (DPDFs) to be extracted.  In this article we criticise the conventional extraction of DPDFs based on `Regge factorisation' in which the exchanged proton is treated as a hadron-like object.  We show using perturbative QCD (pQCD) that the treatment of \emph{diffractive} PDFs has more in common with the \emph{photon} PDFs than with the \emph{proton} PDFs.

\section{Diffractive parton distributions from Regge factorisation} \label{sec:regge}
\begin{figure}
  \centering
  (a)\hspace{0.3\textwidth}(b)\hspace{0.3\textwidth}(c)\\
  \begin{minipage}{0.3\textwidth}
    \includegraphics[width=\textwidth]{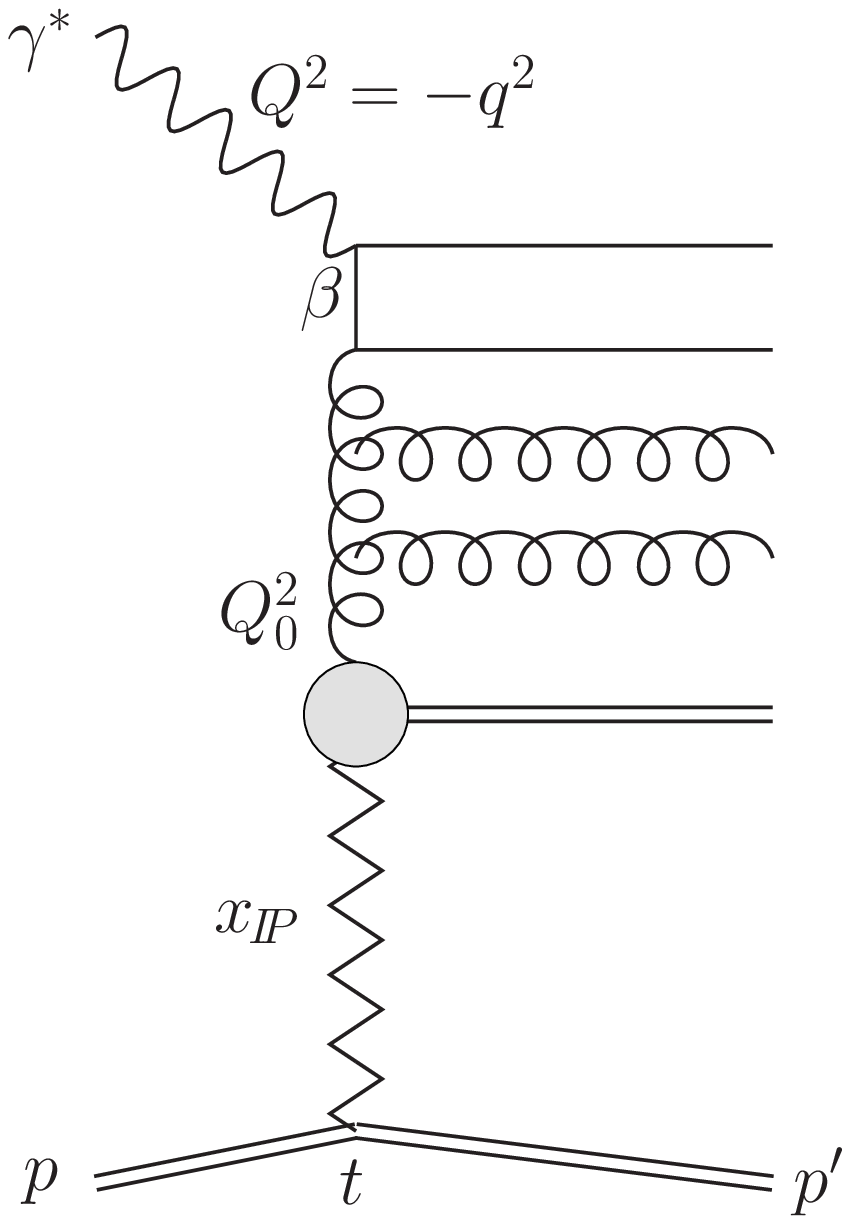}
  \end{minipage}\hfill
  \begin{minipage}{0.3\textwidth}
    \includegraphics[width=\textwidth]{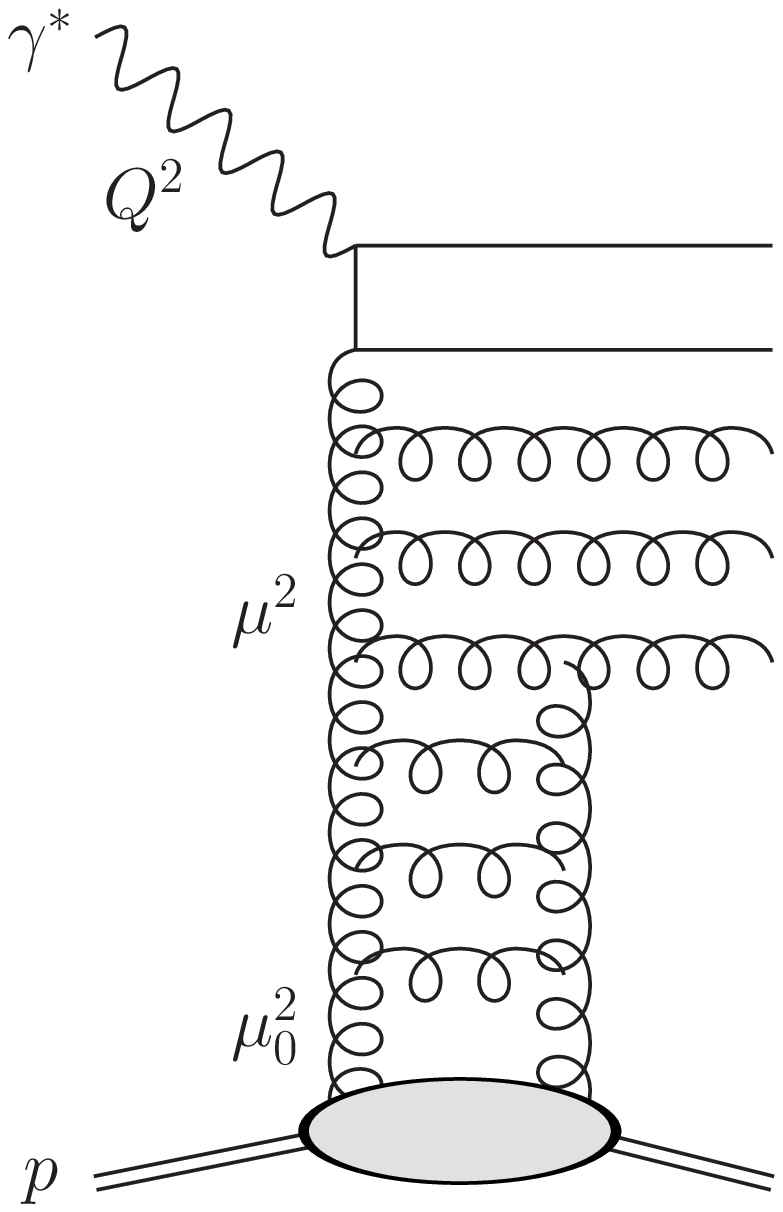}
  \end{minipage}\hfill
  \begin{minipage}{0.3\textwidth}
    \includegraphics[width=\textwidth]{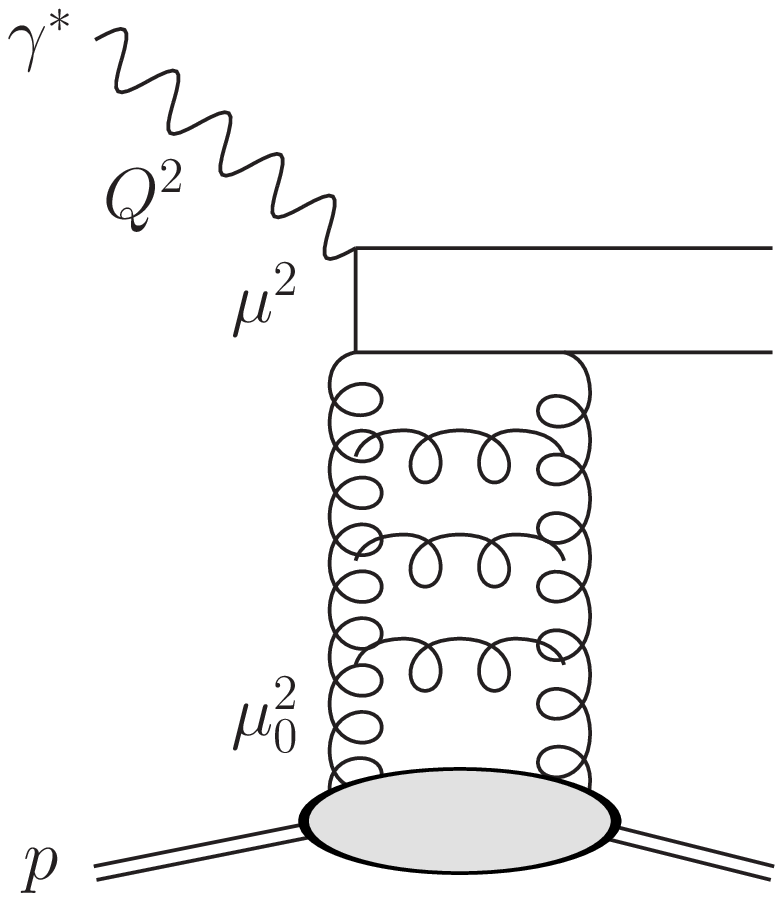}
  \end{minipage}
  \caption{(a) Resolved Pomeron contribution in the `Regge factorisation' approach.  (b) Resolved Pomeron contribution in the `perturbative QCD' approach.  (c) Direct Pomeron contribution in the `perturbative QCD' approach.}
    \label{fig:F2D3}
\end{figure}

Let the momenta of the incoming proton, the outgoing proton, and the photon be labelled $p$, $p^\prime$, and $q$ respectively; see Fig.~\ref{fig:F2D3}(a).  Then the basic kinematic variables in DDIS are the photon virtuality, $Q^2=-q^2$, the Bjorken-$x$ variable, $\xB = Q^2/(2p\cdot q)$, the squared momentum transfer, $t=(p-p^\prime)^2$, the fraction of the proton's light-cone momentum transferred through the rapidity gap, $\xPom=1-{p^\prime}^+/p^+$, and the fraction of the Pomeron's light-cone momentum carried by the struck quark, $\beta=\xB/\xPom$.

It is conventional to extract DPDFs from DDIS data using two levels of factorisation.  Firstly, collinear factorisation means that the diffractive structure function can be written as \cite{Collins:1997sr}
\begin{equation} \label{eq:collfact}
  F_2^{{\rm D}(3)}(\xPom,\beta,Q^2) = \sum_{a=q,g} C_{2,a}\otimes a^{\rm D},
\end{equation}
where the DPDFs $a^{\rm D}=z q^{\rm D}$ or $z g^{\rm D}$, with $z\in[\beta,1]$, satisfy DGLAP evolution:
\begin{equation} \label{eq:DGLAP}
  \frac{\partial a^{\rm D}}{\partial\ln Q^2} = \sum_{a^\prime=q,g}P_{aa^\prime}\otimes{a^\prime}^{\rm D},
\end{equation}
and where $C_{2,a}$ and $P_{aa^\prime}$ are the \emph{same} hard-scattering coefficients and splitting functions as in inclusive DIS.  The factorisation theorem \eqref{eq:collfact} applies when $Q$ is made large, therefore it is correct up to power-suppressed corrections.  It says nothing about the mechanism for diffraction, which is assumed to reside entirely in the input DPDFs fitted to data at a starting scale $Q_0^2$; see Fig.~\ref{fig:F2D3}(a).

In a second stage \cite{Ingelman:1984ns} Regge factorisation is usually assumed, such that
\begin{equation} \label{eq:ReggeFact}
  a^{\rm D}(\xPom,z,Q^2) = f_\Pom(\xPom)\,a^\Pom(z,Q^2),
\end{equation}
where the Pomeron PDFs $a^\Pom=z q^\Pom$ or $z g^\Pom$.  The Pomeron flux factor $f_\Pom$ is taken from Regge phenomenology,
\begin{equation} \label{eq:pomflux}
  f_\Pom(\xPom) = \int_{t_\mathrm{cut}}^{t_\mathrm{min}}\!\dif{t}\quad\mathrm{e}^{B_\Pom\,t}\;\xPom^{1-2\alpha_\Pom(t)}.
\end{equation}
Here, $\alpha_\Pom(t) = \alpha_\Pom(0) + \alpha_\Pom^\prime\,t$, and the parameters $B_\Pom$, $\alpha_\Pom(0)$, and $\alpha_\Pom^\prime$ should be taken from fits to soft hadron data.  Although the first fits to use this approach assumed a `soft' Pomeron, $\alpha_\Pom(0)\simeq1.08$ \cite{Donnachie:1992ny}, all recent fits require a substantially higher value to describe the data.  In addition, a secondary Reggeon contribution is needed to describe the data for $\xPom \gtrsim 0.01$.  This approach is illustrated in Fig.~\ref{fig:F2D3}(a), where the virtualities of the $t$-channel partons are strongly ordered as required by DGLAP evolution.  The Pomeron PDFs $a^\Pom$ are parameterised at some arbitrary low scale $Q_0^2$, then evolved up to the factorisation scale, usually taken to be the photon virtuality $Q^2$.

Although this approach has been found to give a good description of the DDIS data \cite{H1data,Chekanov:2004hy,Abramowicz:2005yc,Newman:2005wm}, it has little theoretical justification.  The `Regge factorisation' of \eqref{eq:ReggeFact} is merely a simple way of parameterising the $x_\Pom$ dependence of the DPDFs.  Note, however, that the effective Pomeron intercept $\alpha_\Pom(0)$ has been observed to depend on $Q^2$ \cite{Chekanov:2005vv}, contrary to the `Regge factorisation' of \eqref{eq:ReggeFact}.  The fact that the required $\alpha_\Pom(0)$ is greater than the `soft' value indicates that there is a significant perturbative QCD (pQCD) contribution to DDIS.

\section{Diffractive parton distributions from perturbative QCD} \label{sec:pQCD}
In pQCD, Pomeron exchange can be described by two-gluon exchange, two gluons being the minimum number needed to reproduce the quantum numbers of the vacuum.  Two-gluon exchange calculations are the basis for the colour dipole model description of DDIS, in which the photon dissociates into $q\bar{q}$ or $q\bar{q}g$ final states.  Such calculations have successfully been used to describe HERA data.  The crucial question, therefore, is how to reconcile two-gluon exchange with collinear factorisation as given by \eqref{eq:collfact} and \eqref{eq:DGLAP}.  Are these two approaches compatible?

\begin{figure}
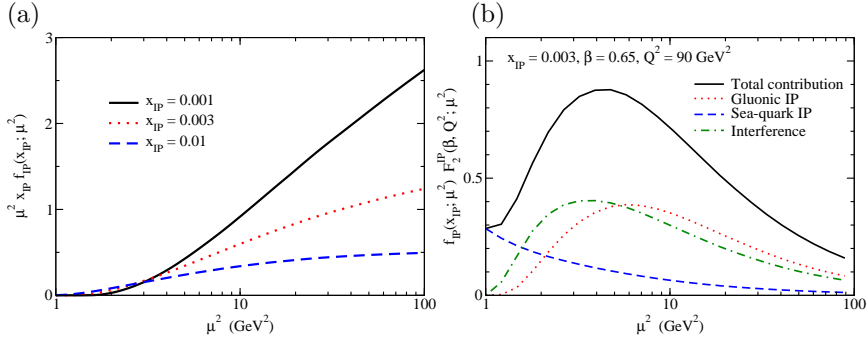

  (a)\hspace{0.5\textwidth}(b)\\
  \includegraphics[clip,width=0.5\textwidth]{watt.fig2a.eps}%
  \includegraphics[clip,width=0.5\textwidth]{watt.fig2b.eps}
  \caption{(a) The perturbative Pomeron flux factor \eqref{eq:fPomG} multiplied by $\mu^2$.  (b) Contributions to $F_{2}^{{\rm D}(3)}$, given by \eqref{eq:f2d3pert}, as a function of $\mu^2$.}
  \label{fig:F2D3pert}
\end{figure}

Generalising the $q\bar{q}$ or $q\bar{q}g$ final states to an arbitrary number of parton emissions from the photon dissociation, and replacing two-gluon exchange by exchange of a parton ladder, we have diagrams like that shown in Fig.~\ref{fig:F2D3}(b) \cite{Ryskin:1990fb,Levin:1992bz,Martin:2004xw,Martin:2005hd}.  Again, the virtualities of the $t$-channel partons are strongly ordered: $\mu_0^2\ll \ldots \ll \mu^2\ll \ldots \ll Q^2$.  The scale $\mu^2$ at which the Pomeron-to-parton splitting occurs can vary between $\mu_0^2\sim 1$ GeV$^2$ and the factorisation scale $Q^2$.  Therefore, to calculate the inclusive diffractive structure function, $F_2^{{\rm D}(3)}$, we need to integrate over $\mu^2$:
\begin{equation} \label{eq:f2d3pert}
  F_{2}^{{\rm D}(3)}(\xPom,\beta,Q^2) = \int_{\mu_0^2}^{Q^2}\!\diff{\mu^2}\;f_{\Pom}(\xPom;\mu^2)\;F_2^\Pom(\beta,Q^2;\mu^2).
\end{equation}
Here, the perturbative Pomeron flux factor can be shown to be \cite{Martin:2005hd}
\begin{equation} \label{eq:fPomG}
  f_{\Pom}(\xPom;\mu^2) = \frac{1}{\xPom B_D} \left[\,R_g\frac{\alpha_S(\mu^2)}{\mu}\;\xPom g(\xPom,\mu^2)\,\right]^2.
\end{equation}
The diffractive slope parameter $B_D$ comes from the $t$-integration, while the factor $R_g$ accounts for the skewedness of the proton gluon distribution \cite{Shuvaev:1999ce}.  There is a similar contribution from sea quarks, where $g(\xPom,\mu^2)$ in \eqref{eq:fPomG} is replaced by $S(\xPom,\mu^2)$, together with an interference term.  In the fits presented here, we use the MRST2001 NLO gluon and sea-quark distributions of the proton \cite{Martin:2001es}.  The Pomeron structure function in \eqref{eq:f2d3pert}, $F_2^\Pom(\beta,Q^2;\mu^2)$, is calculated from Pomeron PDFs, $a^\Pom(z,Q^2;\mu^2)$, evolved using NLO DGLAP from a starting scale $\mu^2$ up to $Q^2$, taking the input distributions to be LO Pomeron-to-parton splitting functions, $a^\Pom(z,\mu^2;\mu^2)=P_{a\Pom}(z)$ \cite{Martin:2004xw,Martin:2005hd}.  At first glance, it would appear that the perturbative Pomeron flux factor \eqref{eq:fPomG} behaves as $f_{\Pom}(\xPom;\mu^2)\sim 1/\mu^2$, so that contributions from large $\mu^2$ are strongly suppressed.  However, at large $\mu^2$, the gluon distribution of the proton behaves as $\xPom g(\xPom,\mu^2)\sim (\mu^2)^\gamma$, where $\gamma$ is the anomalous dimension.  In the BFKL limit of $\xPom\to 0$, $\gamma\simeq 0.5$, so $f_{\Pom}(\xPom;\mu^2)$ would be approximately independent of $\mu^2$.  The HERA domain is in an intermediate region: $\gamma$ is not small, but is less than 0.5.  In Fig.~\ref{fig:F2D3pert}(a) we plot \eqref{eq:fPomG} multiplied by $\mu^2$ to show that \eqref{eq:fPomG} does not behave as $1/\mu^2$ at small $\xPom$.  It is also interesting to plot the integrand of \eqref{eq:f2d3pert} as a function of $\mu^2$, as shown in Fig.~\ref{fig:F2D3pert}(b).  Notice that there is a large contribution from $\mu^2>2$--$3$ GeV$^2$, which is the value of the input scale $Q_0^2$ typically used in the `Regge factorisation' fits of Sect.~\ref{sec:regge}.  Recall that fits using `Regge factorisation' include contributions from $\mu^2\le Q_0^2$ in the input distributions, but neglect all contributions from $\mu^2>Q_0^2$; from Fig.~\ref{fig:F2D3pert}(b) this is clearly an unreasonable assumption.

As well as the \emph{resolved} Pomeron contribution of Fig.~\ref{fig:F2D3}(b), we must also account for the \emph{direct} interaction of the Pomeron in the hard subprocess, Fig.~\ref{fig:F2D3}(c), where there is no DGLAP evolution in the upper part of the diagram.  Therefore, the diffractive structure function can be written as 
\begin{equation}
  F_{2}^{{\rm D}(3)} = \underbrace{\sum_{a=q,g} C_{2,a}\otimes a^{{\rm D}}}_{\text{Resolved Pomeron}} ~+~ \underbrace{C_{2,\Pom}}_{\text{Direct Pomeron}};
\end{equation}
cf.~\eqref{eq:collfact} where there is no direct Pomeron contribution.  The direct Pomeron term, $C_{2,\Pom}$, calculated from Fig.~\ref{fig:F2D3}(c), will again depend on $f_{\Pom}(\xPom;\mu^2)$ given by \eqref{eq:fPomG}.  Therefore, it is formally suppressed by a factor $1/\mu^2$, but in practice does not behave as such; see Fig.~\ref{fig:F2D3pert}(a).

The contribution to the DPDFs from scales $\mu>\mu_0$ is
\begin{equation} \label{eq:apert}
  a^{\rm D}(\xPom,z,Q^2) = \int_{\mu_0^2}^{Q^2}\!\diff{\mu^2}\;f_{\Pom}(\xPom;\mu^2)\;a^\Pom(z,Q^2;\mu^2).
\end{equation}
Differentiating \eqref{eq:apert}, we see that the evolution equations for the DPDFs are \cite{Martin:2005hd}
\begin{equation} \label{eq:evol}
    \frac{\partial a^{\rm D}}{\partial \ln Q^2}
    = \sum_{a^\prime=q,g}P_{aa^\prime}\otimes a^{\prime\,{\rm D}}~+~P_{a\Pom}(z)\,f_\Pom(\xPom;Q^2);
  \end{equation}
cf.~\eqref{eq:DGLAP} where the second term of \eqref{eq:evol} is absent.  That is, the DPDFs satisfy an \emph{inhomogeneous} evolution equation \cite{Levin:1992bz,Martin:2005hd}, with the extra inhomogeneous term in \eqref{eq:evol} leading to more rapid evolution than in the `Regge factorisation' fits described in Sect.~\ref{sec:regge}.  Note that the inhomogeneous term will change the $\xPom$ dependence evolving upwards in $Q^2$, in accordance with the data, and unlike the `Regge factorisation' assumption \eqref{eq:ReggeFact}.  Again, the inhomogeneous term in \eqref{eq:evol} is formally suppressed by a factor $1/Q^2$, but in practice does not behave as such; see Fig.~\ref{fig:F2D3pert}(a).

\begin{figure}
  \centering
  \begin{tabular}{c|c|c}
    & Resolved photon & Direct photon \\
    & ($x_\gamma < 1$) & ($x_\gamma = 1$) \\ \hline
    \begin{minipage}{0.2\textwidth}\centering Resolved Pomeron \\ ($z_\Pom < 1$)\end{minipage} & \begin{minipage}{0.25\textwidth}\includegraphics[width=\textwidth,clip]{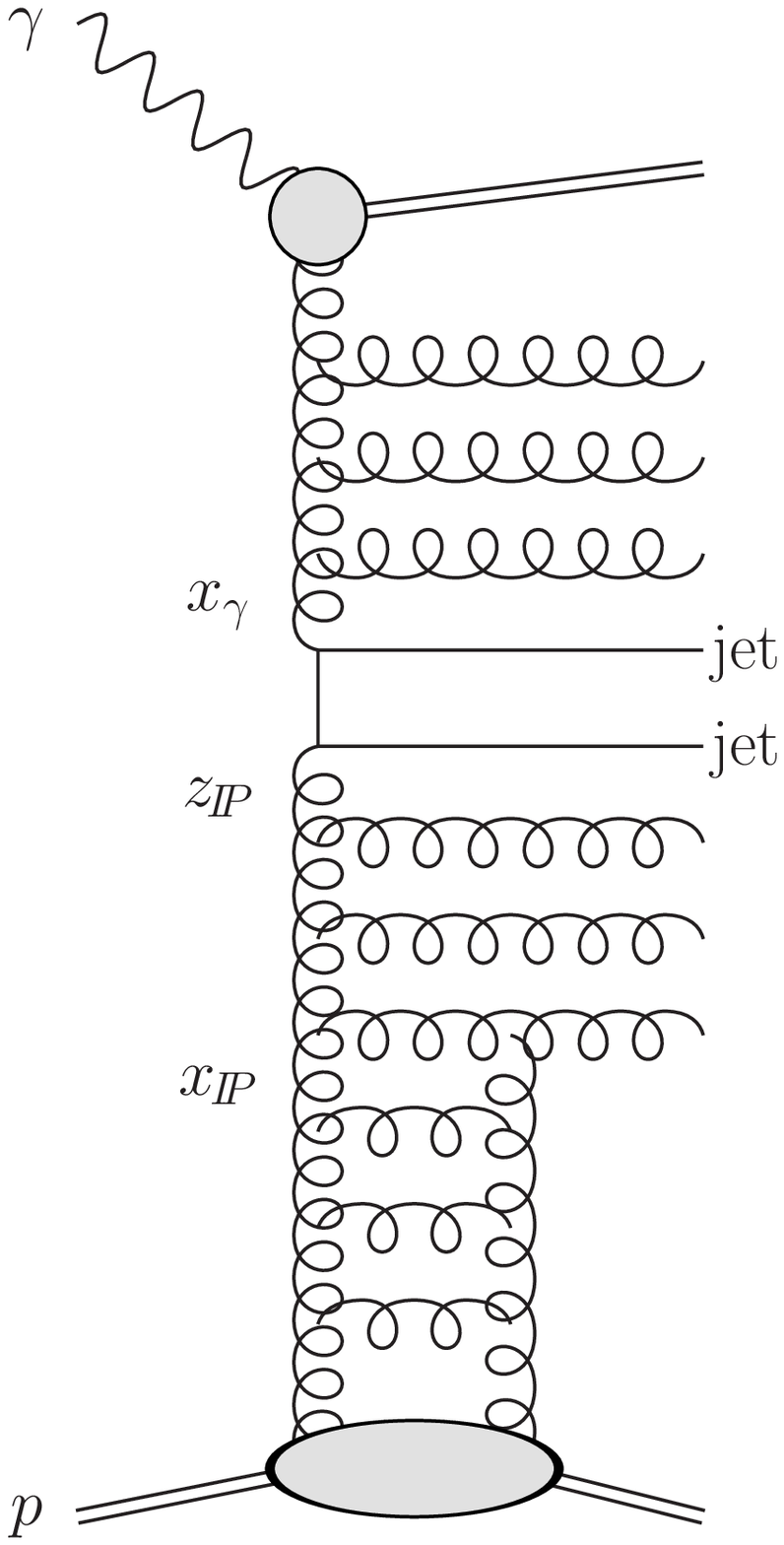}\end{minipage} & \begin{minipage}{0.25\textwidth}\includegraphics[width=\textwidth,clip]{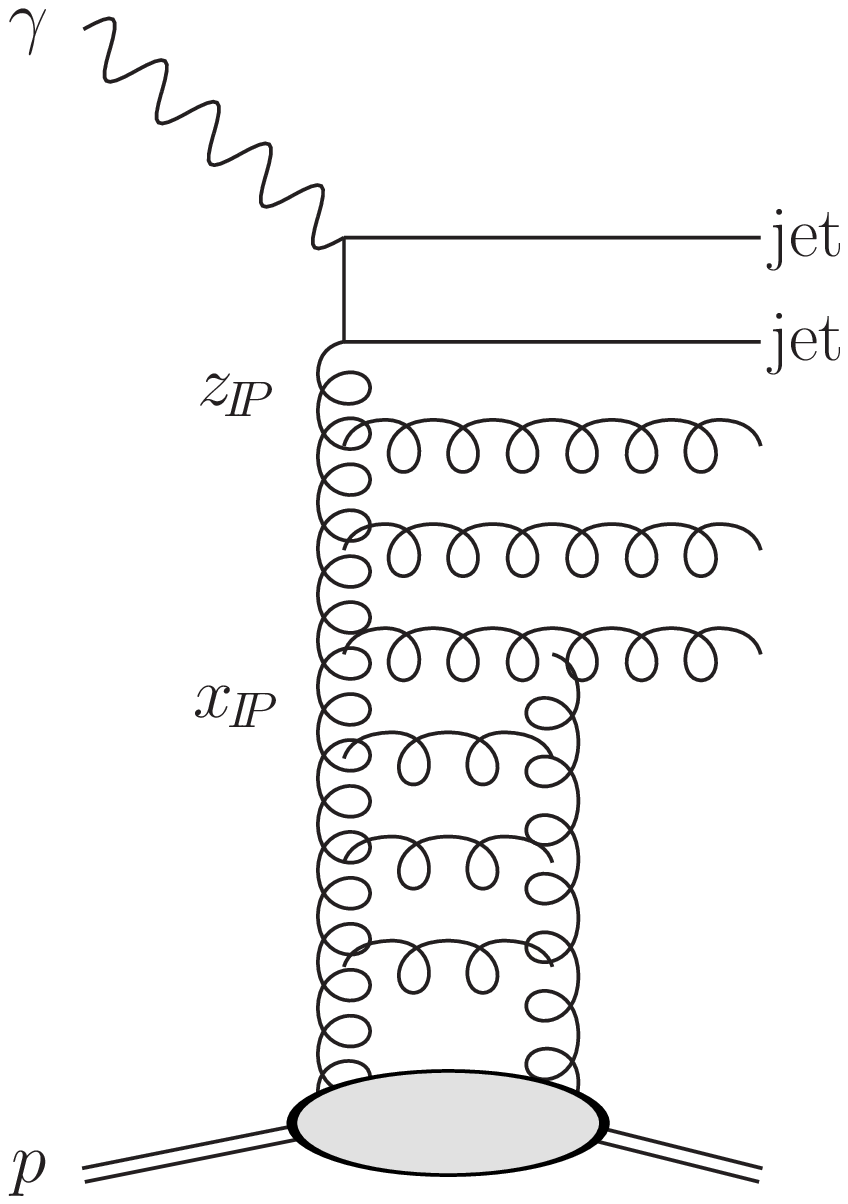} \end{minipage} \\ \hline
    \begin{minipage}{0.2\textwidth}\centering Direct Pomeron \\ ($z_\Pom = 1$) \end{minipage} & \begin{minipage}{0.25\textwidth}\includegraphics[width=\textwidth,clip]{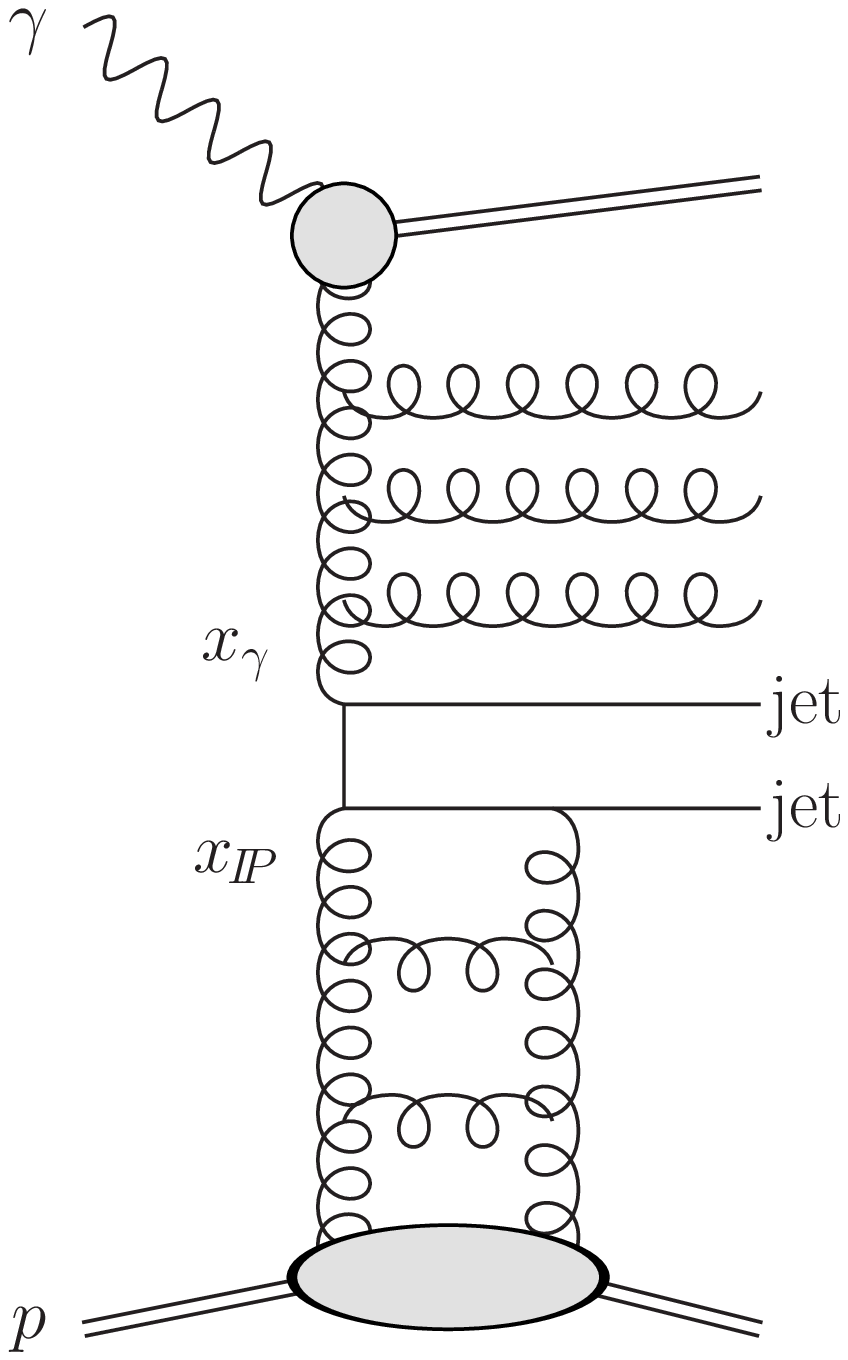}\end{minipage} & \begin{minipage}{0.25\textwidth}\includegraphics[width=\textwidth,clip]{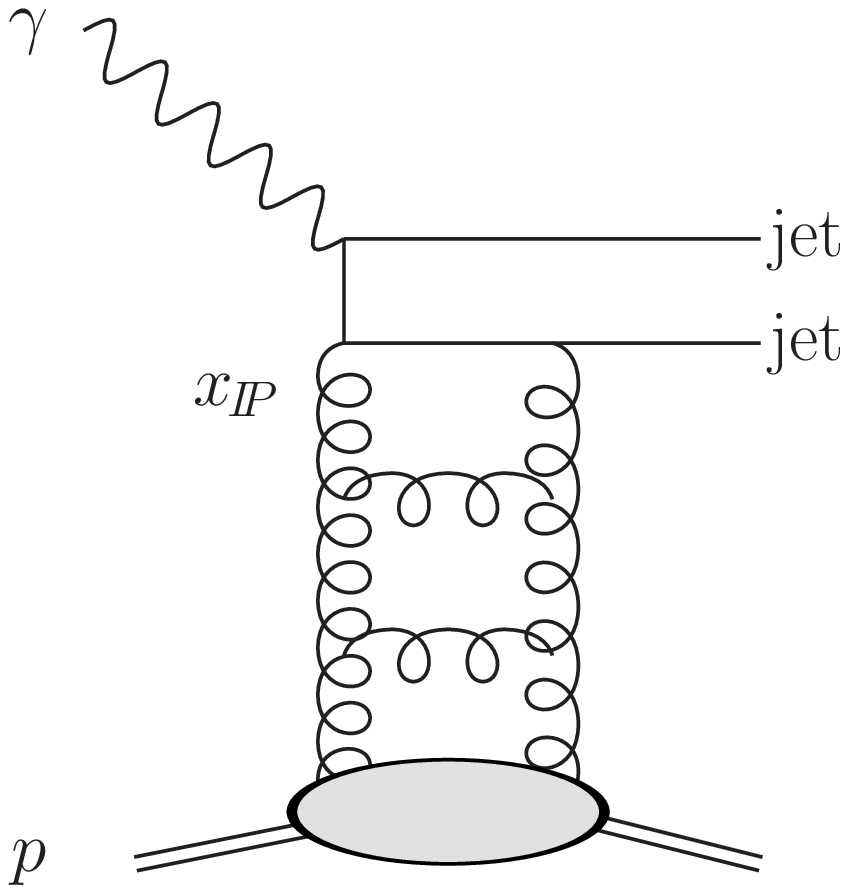}\end{minipage}
  \end{tabular}
  \caption{The four classes of contributions to diffractive dijet photoproduction at LO.  Both the photon and the Pomeron can be either `resolved' or `direct'.}
  \label{fig:dijet}
\end{figure}

Therefore, the diffractive structure function is analogous to the photon structure function, where there are both resolved and direct components and the photon PDFs satisfy an inhomogeneous evolution equation, where at LO the inhomogeneous term accounts for the splitting of the point-like photon into a $q\bar{q}$ pair.  If we consider, for example, diffractive dijet photoproduction, there are four classes of contributions; see Fig.~\ref{fig:dijet}.  The relative importance of each contribution will depend on the values of $x_\gamma$, the fraction of the photon's momentum carried by the parton entering the hard subprocess, and $z_{\Pom}$, the fraction of the Pomeron's momentum carried by the parton entering the hard subprocess.

\section{Description of DDIS data}

A NLO analysis of DDIS data is not yet possible.  The direct Pomeron terms, $C_{2,\Pom}$, and Pomeron-to-parton splitting functions, $P_{a\Pom}$, need to be calculated at NLO within a given factorisation scheme (for example, $\overline{\textrm{MS}}$).  Here, we perform a simplified analysis where the usual coefficient functions $C_{2,a}$ and splitting functions $P_{aa^\prime}$ ($a,a^\prime=q,g$) are taken at NLO, but $C_{2,\Pom}$ and $P_{a\Pom}$ are taken at LO \cite{Martin:2005hd}.  We work in the fixed flavour number scheme, where there is no charm DPDF.  Charm quarks are produced via $\gamma^* g^\Pom\to c\bar{c}$ at NLO \cite{Riemersma:1994hv} and $\gamma^*\Pom\to c\bar{c}$ at LO \cite{Levin:1996vf}.  For light quarks, we include the direct Pomeron process $\gamma^*_L\Pom\to q\bar{q}$ at LO \cite{Martin:2005hd}, which is higher-twist and known to be important at large $\beta$.

To see the effect of the direct Pomeron contribution and the inhomogeneous evolution, we make two types of fits:
\begin{itemize}
  \item[``Regge''] : The `Regge factorisation' approach discussed in Sect.~\ref{sec:regge}, where there is no direct Pomeron contribution and no inhomogeneous term in the evolution equation.
  \item[``pQCD''] : The `perturbative QCD' approach discussed in Sect.~\ref{sec:pQCD}, where these effects are included.
\end{itemize}

\begin{figure}
  (a)\hspace{0.5\textwidth}(b)\\
  \includegraphics[width=0.5\textwidth]{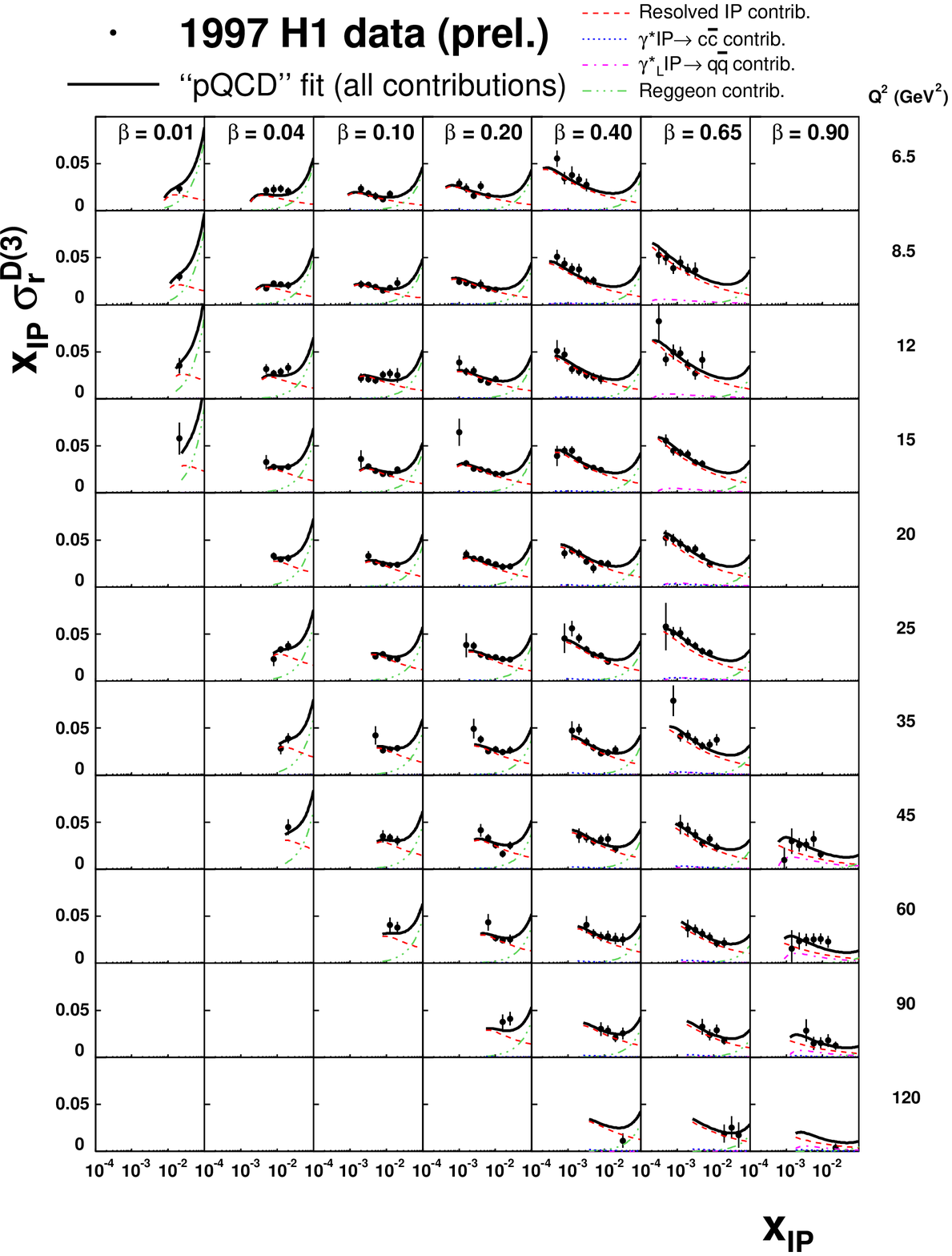}%
  \includegraphics[width=0.5\textwidth]{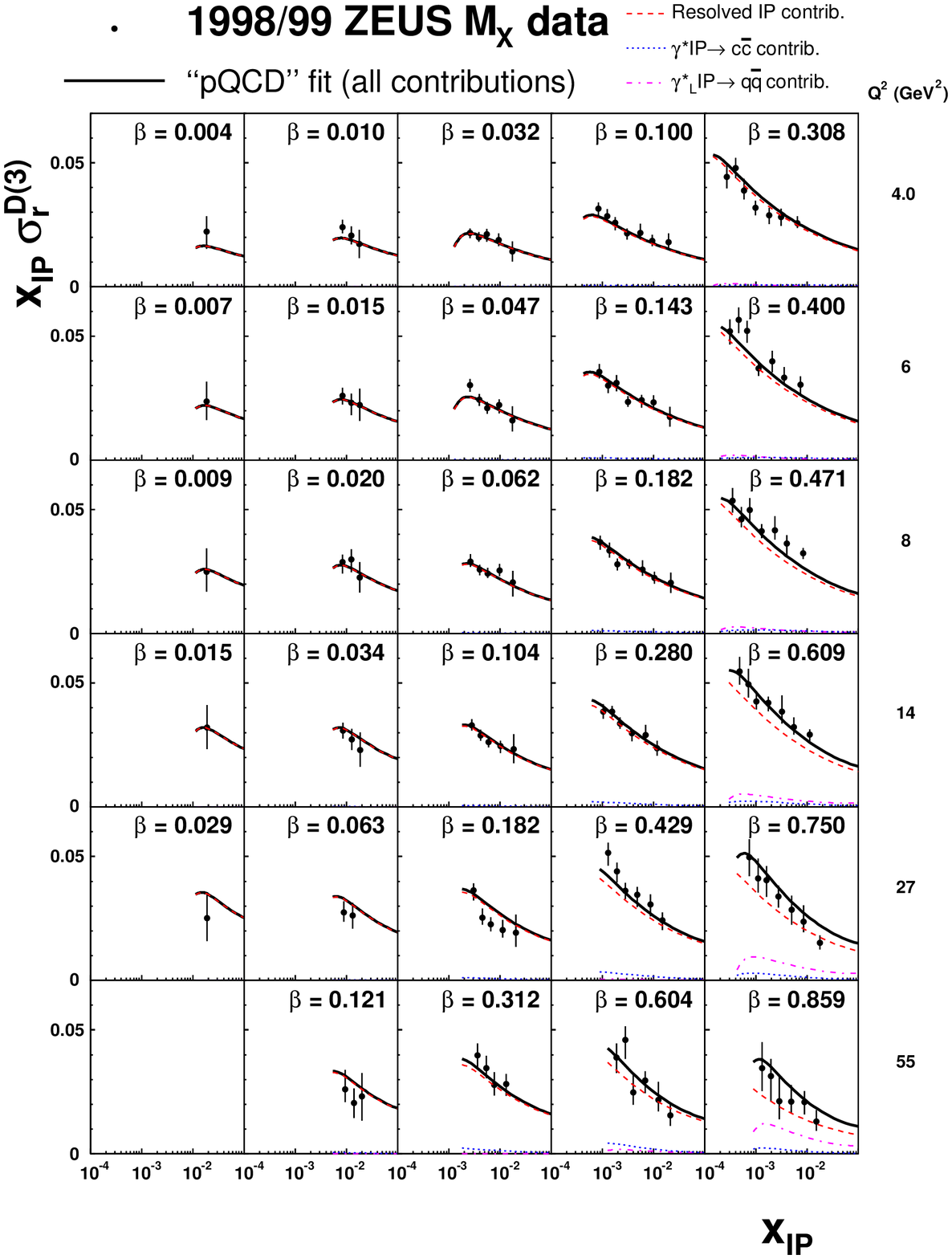}
  \caption{``pQCD'' fits to (a) H1 LRG and (b) ZEUS $M_X$ data.}
  \label{fig:data}
\end{figure}

\begin{figure}
  \centering
  \includegraphics[width=\textwidth,clip]{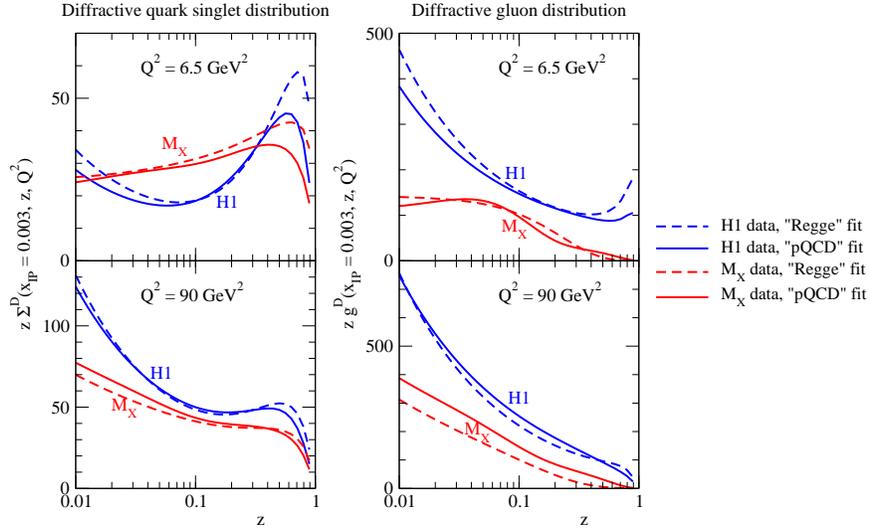}
  \caption{DPDFs obtained from separate fits to H1 LRG and ZEUS $M_X$ data using the ``Regge'' and ``pQCD'' approaches.}
  \label{fig:dpdfs}
\end{figure}

We make separate fits to the recent H1 LRG (prel.) \cite{H1data} and ZEUS $M_X$ \cite{Chekanov:2005vv} $\sigma_r^{{\rm D}(3)}$ data, applying cuts $Q^2\ge3$ GeV$^2$ and $M_X\ge2$ GeV, and allowing for overall normalisation factors of $1.10$ and $1.43$ to account for proton dissociation up to masses of $1.6$ GeV and $2.3$ GeV respectively.  Statistical and systematic experimental errors are added in quadrature.  The strong coupling is set via $\alpha_S(M_Z)=0.1190$.  We take the input forms of the DPDFs at a scale $Q_0^2=3$ GeV$^2$ to be
\begin{align}
  z\Sigma^{\rm D}(\xPom,z,Q_0^2) &= f_\Pom(\xPom)\; C_q\,z^{A_q}(1-z)^{B_q},\label{eq:inputq} \\
  z g^{\rm D}(\xPom,z,Q_0^2) &= f_\Pom(\xPom)\; C_g\,z^{A_g}(1-z)^{B_g},\label{eq:inputg}
\end{align}
where $f_\Pom(\xPom)$ is given by \eqref{eq:pomflux}, and where $\alpha_{\Pom}(0)$, $C_a$, $A_a$, and $B_a$ ($a=q,g$) are free parameters.  The secondary Reggeon contribution to the H1 data is treated in a similar way as in the H1 2002 fit \cite{H1data}, using the GRV pionic parton distributions \cite{Gluck:1991ey}.  Good fits are obtained in all cases, with $\chisq=0.75$, $0.71$, $0.76$, and $0.84$ for the ``Regge'' fit to H1 data, ``pQCD'' fit to H1 data, ``Regge'' fit to ZEUS $M_X$ data, and ``pQCD'' fit to ZEUS $M_X$ data respectively.  The ``pQCD'' fits are shown in Fig.~\ref{fig:data}, including a breakdown of the different contributions.  The DPDFs are shown in Fig.~\ref{fig:dpdfs}.
Note that the ``pQCD'' DPDFs are smaller than the corresponding ``Regge'' DPDFs at large $z$ due to the inclusion of the higher-twist $\gamma_L^*\Pom\to q\bar{q}$ contribution.  Also note that the ``pQCD'' DPDFs have slightly more rapid evolution than the ``Regge'' DPDFs due to the extra inhomogeneous term in the evolution equation \eqref{eq:evol}.  There is a large difference between the DPDFs obtained from the H1 LRG and ZEUS $M_X$ data due to the different $Q^2$ dependence of these data sets; see also \cite{Abramowicz:2005yc,Newman:2005wm}.

\begin{figure}
  (a)\hspace{0.5\textwidth}(b)\\
  \includegraphics[width=0.5\textwidth,clip]{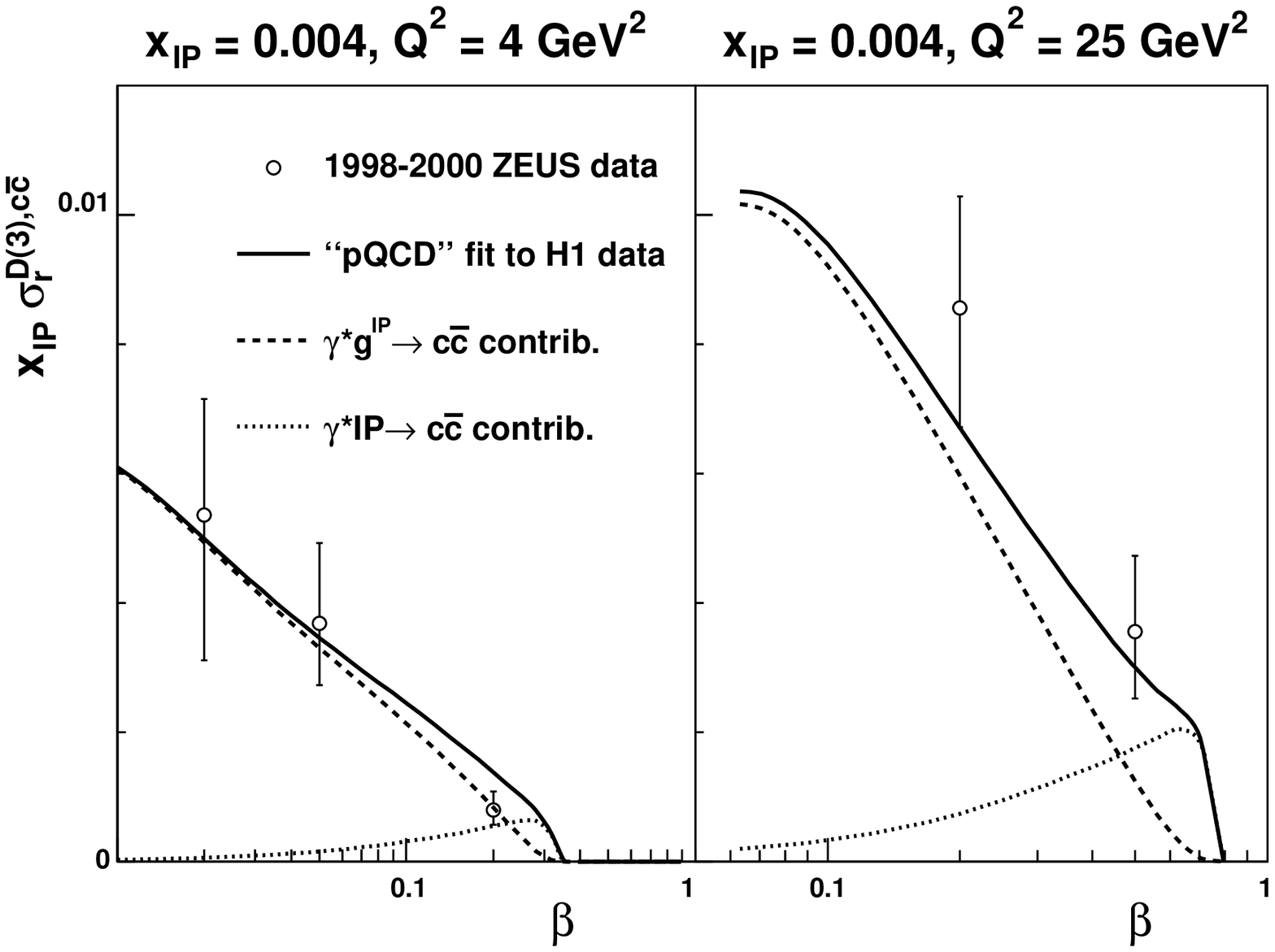}%
  \includegraphics[width=0.5\textwidth,clip]{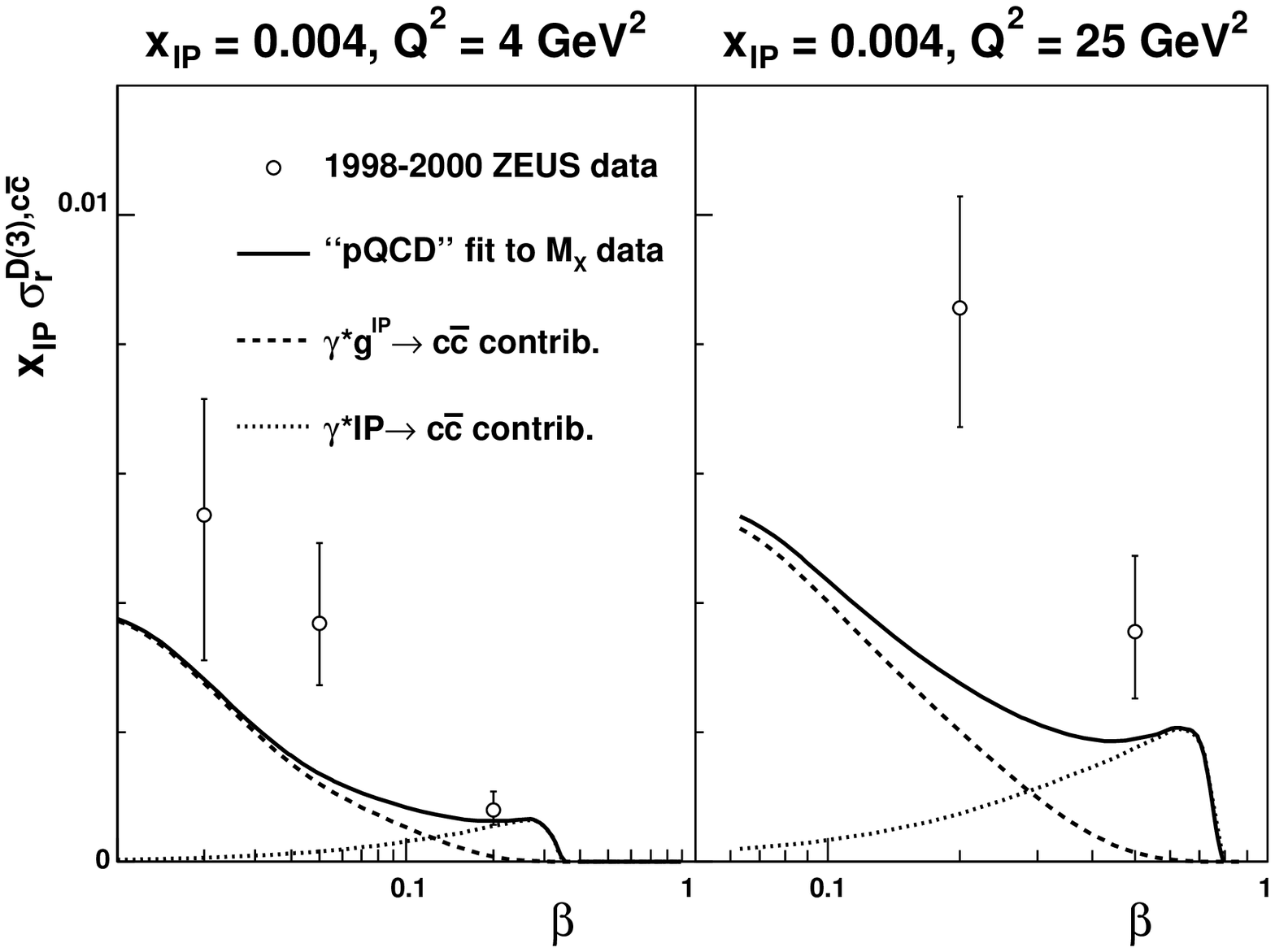}
  \caption{Predictions for ZEUS LRG diffractive charm production data using DPDFs from the ``pQCD'' fits to (a) H1 LRG and (b) ZEUS $M_X$ data.  Note the large direct Pomeron ($\gamma^* \Pom\to c\bar{c}$) contribution at moderate $\beta$.}
  \label{fig:charm}
\end{figure}

The predictions from the two ``pQCD'' fits for the charm contribution to the diffractive structure function as measured by ZEUS using the LRG method \cite{Chekanov:2003gt} are shown in Fig.~\ref{fig:charm}.  Our H1 LRG fit gives a good description, while our ZEUS $M_X$ fit is too small at low $\beta$.  Note that the direct Pomeron contribution is significant at moderate $\beta$.  These charm data points were included in the determination of DPDFs from ZEUS LPS data \cite{Chekanov:2004hy}, but only the resolved Pomeron ($\gamma^* g^\Pom\to c\bar{c}$) contribution was included and not the direct Pomeron ($\gamma^* \Pom\to c\bar{c}$) contribution.  Therefore, the diffractive gluon distribution from the ZEUS LPS fit \cite{Chekanov:2004hy} needed to be artificially large to fit the charm data at moderate $\beta$.

\section{Conclusions and outlook}

To summarise, diffractive DIS is more complicated to analyse than inclusive DIS.  Collinear factorisation holds, but we need to account for the direct Pomeron coupling, leading to an inhomogeneous evolution equation \eqref{eq:evol}.\footnote{The inhomogeneous evolution of DPDFs leads, via the AGK cutting rules \cite{Abramovsky:1973fm}, to non-linear evolution of the inclusive PDFs \cite{Martin:2004xx,Watt:2005iu}.}  Therefore, the treatment of DPDFs has more in common with photon PDFs than with proton PDFs.  The H1 LRG and ZEUS $M_X$ data have a different $Q^2$ dependence, leading to different DPDFs.  This issue needs further attention.  For a NLO analysis of DDIS data, the direct Pomeron terms, $C_{2,\Pom}$, and Pomeron-to-parton splitting functions, $P_{a\Pom}$, need to be calculated at NLO.  There are indications \cite{Levin:1996vf} that there are large $\pi^2$-enhanced virtual loop corrections (`K-factors') similar to those found in the Drell--Yan process.  As with all PDF determinations, the sensitivity to the form of the input parameterisation, \eqref{eq:inputq} and \eqref{eq:inputg}, and input scale $Q_0^2$ needs to be studied.  The inclusion of jet and heavy quark DDIS data, and possibly $F_L^{{\rm D}(3)}$ if it is measured \cite{Newman:2005mm}, would help to constrain the DPDFs further.  The extraction of DPDFs from HERA data will provide an important input for predictions of diffractive processes at the LHC.


\end{document}